\documentclass[sigconf]{acmart}

\usepackage{soul}       
\usepackage{subcaption} 
\usepackage{multirow}   
\usepackage{colortbl}   
\usepackage{ifthen}     
\usepackage{tikz}
\usepackage{balance}
\AtBeginDocument{%
  }

\setcopyright{acmlicensed}
\copyrightyear{2018}
\acmYear{2018}
\acmDOI{XXXXXXX.XXXXXXX}
\acmConference[Conference acronym 'XX]{Make sure to enter the correct
  conference title from your rights confirmation email}{June 03--05,
  2018}{Woodstock, NY}
\acmISBN{978-1-4503-XXXX-X/2018/06}

\copyrightyear{2026}
\acmYear{2026}
\setcopyright{cc}
\setcctype{by-nc-nd}
\acmConference[SIGIR-AP 2026]{Proceedings of the 2026 Annual International ACM SIGIR Conference on Research and Development in Information Retrieval in the Asia Pacific Region}{December 13--15, 2026}{Singapore, Singapore}
\acmBooktitle{Proceedings of the 2026 Annual International ACM SIGIR Conference on Research and Development in Information Retrieval in the Asia Pacific Region (SIGIR-AP 2026), December 13--15, 2026, Singapore, Singapore}
\acmDOI{10.1145/3848000.3850197}
\acmISBN{979-8-4007-3033-7/2026/12}

\begin{document}

\title{VisR-RAG-Fin: Benchmarking Multimodal RAG over Visually Rich Financial Documents}


\author{Omar El Bachyr}
\affiliation{%
  \institution{University of Luxembourg}
  \country{Luxembourg}
}
\email{omar.elbachyr@uni.lu}

\author{Laura Maria Bernardy}
\affiliation{%
  \institution{University of Luxembourg}
  \country{Luxembourg}
}
\email{laura.bernardy@uni.lu}

\author{Saad Ezzini}
\affiliation{%
  \institution{King Fahd University of Petroleum and Minerals}
  \country{Saudi Arabia}
}
\email{saad.ezzini@kfupm.edu.sa}

\author{Jacques Klein}
\affiliation{%
  \institution{University of Luxembourg}
  \country{Luxembourg}
}
\email{jacques.klein@uni.lu}

\author{Tegawendé F. Bissyandé}
\affiliation{%
  \institution{University of Luxembourg}
  \country{Luxembourg}
}
\email{tegawende.bissyande@uni.lu}

\author{Charles Monté}
\affiliation{%
  \institution{Banque BGL BNP Paribas}
  \country{Luxembourg}
}
\email{charles.monte@bgl.lu}

\author{Anas Zilali}
\affiliation{%
  \institution{Banque BGL BNP Paribas}
  \country{Luxembourg}
}
\email{anas.zilali@bgl.lu}

\author{Ulrick Ble}
\affiliation{%
  \institution{Banque BGL BNP Paribas}
  \country{Luxembourg}
}
\email{ulrick.ble@bgl.lu}

\author{Anne Goujon}
\affiliation{%
  \institution{Banque BGL BNP Paribas}
  \country{Luxembourg}
}
\email{anne.goujon@bgl.lu}

\renewcommand{\shortauthors}{El Bachyr et al.}

\begin{abstract} 
Enterprise documents such as financial reports are increasingly processed by automated systems, yet they are primarily designed for human reading and exhibit complex visual structure, combining dense text with tables, charts, diagrams, and rich layouts. These characteristics pose significant challenges for Retrieval-Augmented Generation (RAG) systems, which must retrieve relevant pages and leverage visual and structural information during answer generation. Despite recent advances in multimodal retrieval and vision–language models, there remains a limited understanding of how design choices across retrieval and generation components affect end-to-end RAG performance on visually rich documents. In this paper, we present a comprehensive empirical study of multimodal RAG for page-level Question Answering (QA) over financial documents. We introduce \textit{VisR-RAG-Fin (Visually Rich RAG for Finance)}, an open-source benchmark constructed through a general pipeline for visually rich documents that preserves original layout and visual content. Using this benchmark, we systematically analyze retrieval and generation behavior under text-only, multimodal, and hybrid configurations, providing a foundation for rigorous evaluation of RAG systems on complex enterprise document collections. Overall, this work provides actionable insights for practitioners designing and evaluating multimodal RAG pipelines over complex, visually rich enterprise documents.
\end{abstract}





\maketitle

\section{Introduction}
\raggedbottom
Retrieval-Augmented Generation (RAG) systems are increasingly adopted in industrial settings to support Question Answering over large collections of enterprise documents. In domains such as finance, documents are often lengthy and visually complex, where dense text coexists with tables, charts, diagrams, and complex layouts. This makes it challenging not only to retrieve the correct document pages but also to effectively exploit the visual and structural information within the retrieved pages to produce accurate answers.
Despite its practical relevance, the evaluation of RAG pipelines remains largely text-centric~\citep{yu2024evaluation}. Most benchmarks and experimental setups rely on plain-text corpora or flattened document representations, thereby removing layout and visual cues that are essential in real-world documents. Consequently, these evaluations provide limited insight into retrieval and generation performance on visually structured documents encountered in real-world RAG deployments, such as those common in the financial domain.

Recent advances in multimodal retrieval~\citep{lee2024unified, faysse2024colpali} and Vision~Language Models (VLMs) indicate that visual features can benefit document understanding tasks. However, important questions remain regarding their role in realistic RAG settings. In particular, differences between text-only and visual retrieval methods on visually rich enterprise documents have not been fully examined. Moreover, the effectiveness of multimodal retrieval across different evidence types, such as narrative text, tables, and visually grounded elements, remains insufficiently examined. Beyond retrieval, overall system effectiveness also depends on how retrieved content is represented and consumed by the generation model. In practice, retrieved pages can be provided as text, images, or a combination of both, depending on system design and model capabilities. These representation choices influence how effectively the generation model can interpret and integrate the retrieved evidence, yet their impact on end-to-end RAG performance and robustness has not been systematically studied.
To address these gaps, we introduce \textit{VisR-RAG-Fin}, a benchmark designed for realistic evaluation of Retrieval-Augmented Question Answering over visually rich financial documents. VisR-RAG-Fin operates at the page level and preserves the original visual structure of documents, enabling controlled evaluation of both retrieval and generation components in multimodal RAG pipelines. The benchmark is constructed through a \textit{human-in-the-loop} annotation process, in which sampled visually rich pages are manually verified, while generated question–answer (QA) pairs are manually reviewed and annotated by human annotators to ensure correctness and relevance for page-level question answering over financial documents.
Using VisR-RAG-Fin, we conduct an empirical study of multimodal RAG systems. On the retrieval side, we compare textual retrievers with multimodal retrievers, including both single-vector and multi-vector multimodal models, and examine whether multimodal retrievers leveraging visual features offer advantages over textual models for queries whose answers rely on visual or layout-dependent evidence, such as tables, diagrams, or images. We additionally evaluate the effect of hard negative page mining, contrasting it with random negative sampling over non-relevant pages, to increase the difficulty of the retrieval benchmark. For the generation side, we evaluate answer correctness in an end-to-end multimodal RAG pipeline by contrasting oracle page access with retrieval-based context. Oracle access provides an upper bound on generation performance, while retrieval-based settings assess generation behavior under realistic conditions as the number of retrieved document pages provided as context varies. Building on this analysis, we further compare text-only, multimodal, and hybrid RAG pipelines, in which retrieved context is provided as text, images, or a mixture of both. Together, these experiments provide a detailed view of how retrieval and generation design choices interact in practice, highlighting limitations that are particularly relevant for the deployment of RAG systems over visually rich enterprise documents, with financial documents from our VisR-RAG-Fin benchmark serving as a representative class of such real-world enterprise documents. 

Our contributions are summarized as follows:
\noindent\textbf{(1) Benchmark construction methodology.} A general, reusable pipeline for building realistic retrieval-augmented QA benchmarks over visually rich documents, applicable across industrial domains.
\noindent\textbf{(2) VisR-RAG-Fin benchmark.} An open-source benchmark for page-level retrieval and Question Answering over visually rich financial documents, preserving original visual and layout information, and constructed using the proposed pipeline.
\noindent\textbf{(3) Empirical analysis of multimodal RAG.} Leveraging the proposed benchmark, we benchmark state-of-the-art retrieval and generation models to analyze (i) retrieval effectiveness across textual and multimodal retrievers, (ii) the impact of visual evidence and context modality on answer correctness in end-to-end RAG pipelines, and (iii) the gap between oracle and retrieval-based generation.

By jointly analyzing retrieval and generation under realistic conditions, our analysis provides actionable insights for designing and evaluating multimodal RAG systems in industrial document settings. The benchmark, annotation guidelines, prompts, and evaluation code are available in our \href{https://github.com/OmarElbachyr/VisR-RAG-Fin}{\textbf{GitHub repository}}.

\section{Related Work}
\raggedbottom
\noindent{\textbf{Document Visual Question Answering (DocVQA).}} DocVQA is considered a subtask of Document Understanding (DU) and focuses on answering questions over document images by integrating textual, layout, and visual information. Earlier benchmarks introduced single-page datasets such as DocVQA~\citep{mathew2021docvqa}, which provide extractive QA over scanned industrial documents, VisualMRC~\citep{tanaka2021visualmrc} using open-domain questions on webpage screenshots, InfoVQA~\citep{mathew2022infographicvqa} with reasoning over visually rich infographics, and TAT-DQA~\citep{zhu2022towards} focused on discrete reasoning (e.g., arithmetic and comparison operations on numbers) in financial reports. Then, to mitigate single-page setup limitations, later benchmarks introduced multi-page and long-context challenges: DUDE~\citep{van2023document} offered diverse multi-domain, multi-layout, and multi-answer-type questions, MP-DocVQA~\citep{tito2023hierarchical} extended single-page tasks of DocVQA by including adjacent pages, and SlideVQA~\citep{tanaka2023slidevqa} introduced reasoning across slides with numerical and multi-hop QA. More recent benchmarks like MMLongBench-Doc~\citep{ma2024mmlongbench} and DocBench~\citep{zou2024docbench} have redefined the DocVQA task by introducing longer context settings that require models to reason over entire documents that span multiple pages. While some of these benchmarks also involve retrieval, VisR-RAG-Fin specifically focuses on evaluating page-level retrieval and end-to-end RAG over visually rich financial documents, incorporating evidence-type annotations and challenging within-document hard negatives.

\noindent{\textbf{Multimodal Retrievers.}}
Recent work~\citep{ma2024unifying, faysse2024colpali, yu2024visrag} in information retrieval has moved beyond traditional text-only systems by directly integrating visual and layout information. Work in~\citep{ma2024unifying} proposes encoding full document screenshots using Vision–Language Models, preserving both textual and visual cues without relying on OCR. Similarly, ColPali~\citep{faysse2024colpali} exposes the limitations of text-centric retrieval for visually rich documents and adapts the ColBERT~\citep{khattab2020colbert} late-interaction paradigm to page images, generating multi-vector visual embeddings that enable fine-grained matching between query tokens and localized visual regions within document pages, showing substantial improvements in retrieval. Building on this foundation, subsequent work~\citep{nomicembedmultimodal2025, xu2025llama} has proposed stronger late-interaction multimodal retrievers that retain the ColPali design while leveraging more capable VLM backbones, resulting in improved retrieval effectiveness.

\noindent{\textbf{Multimodal RAG.}}
Recent work in document Question Answering has moved beyond text-only Retrieval-Augmented Generation~\cite{lewis2020retrieval} toward multimodal RAG systems that combine text with images and layout information. Early approaches such as VisRAG~\citep{yu2024visrag} rely on page-level visual embeddings to preserve document structure that is often lost during text extraction. Benchmarks such as MMDocRAG~\citep{dong2025benchmarking} and UniDoc-Bench~\citep{peng2025unidoc} investigate multimodal evidence retrieval and fusion. Complementing these efforts, VisR-RAG-Fin focuses on financial PDFs, combining evidence-type annotations and within-document hard negatives with the evaluation of both page-level retrieval and downstream answer generation.

\section{VisR-RAG-Fin Benchmark Construction}
\begin{figure}[t]
    \centering
    \scalebox{0.5}{
        \includegraphics[width=\textwidth]{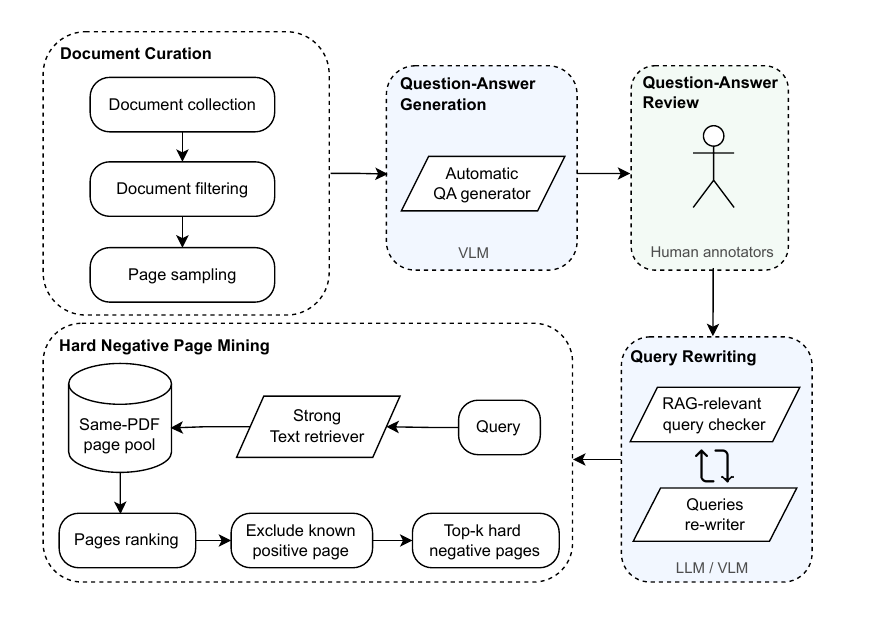}
    }
    \caption{End-to-end construction pipeline of the \textit{VisR-RAG-Fin} benchmark.}
    \label{fig:benchmark_construction_pipeline}
\end{figure}

\raggedbottom
The \textit{VisR-RAG-Fin} benchmark is built using a multi-stage construction pipeline, shown in Figure~\ref{fig:benchmark_construction_pipeline}, to ensure quality, diversity, and suitability for RAG evaluation. The process comprises seven successive steps: (1) document collection, (2) document filtering, (3) page sampling, (4) Question-Answer generation, (5) Question-Answer review, (6) query rewriting, and (7) hard negative page mining. Each step progressively refines the data, from raw document acquisition to the creation of high-quality queries that form a challenging Retrieval-Augmented Question Answering benchmark.

\subsection{Document Collection}
To construct a benchmark for visually rich query–page retrieval in the financial domain, we curated a collection of documents that were publicly published by nine major European banking institutions (Table~\ref{tab:en_collected_docs}). These institutions were selected through an exploratory web-based survey targeting large European banks with publicly available financial reports. We focused on document types such as annual reports, regulatory filings, and financial statements, which are inherently visually rich and characterized by a heterogeneous mixture of text, tables, figures, and layout-dependent structures. As financial documents are distributed across diverse platforms with varying degrees of accessibility, we adopted two complementary document collection strategies to jointly maximize corpus scale and coverage.

\begin{table}[t]
\centering
\caption{Statistics of collected English-language PDF documents by bank (Total = 791 PDFs). Page-level statistics report the total number of pages and the mean number of pages per document.}
\label{tab:en_collected_docs}
\tiny
\setlength{\tabcolsep}{1.2pt}
\renewcommand{\arraystretch}{1.05}
\begin{tabular}{@{}lrrrrrrrrr@{}}
\toprule
\textbf{Statistic}
& \textbf{BPCE}
& \textbf{Santander}
& \textbf{UBS}
& \textbf{Barclays}
& \shortstack{\textbf{Société}\\\textbf{Générale}}
& \shortstack{\textbf{BGL BNP}\\\textbf{Paribas}}
& \shortstack{\textbf{Deutsche}\\\textbf{Bank}}
& \shortstack{\textbf{Crédit}\\\textbf{Agricole}}
& \shortstack{\textbf{Lloyds}\\\textbf{BG}} \\
\midrule
\textbf{PDFs}
& 162 & 153 & 114 & 100 & 73 & 62 & 57 & 37 & 33 \\
\textbf{Total pages}
& 7,499 & 2,842 & 7,488 & 32,394 & 10,519 & 9,871 & 1,606 & 1,180 & 5,659 \\
\textbf{Mean pages}
& 46.3 & 18.6 & 65.7 & 323.9 & 144.1 & 159.2 & 28.2 & 31.9 & 171.5 \\
\bottomrule
\end{tabular}
\end{table}

\noindent\textbf{Search Engine-Driven Document Collection.} We use domain- and file-type–specific Google queries to identify relevant documents hosted on institutional websites. For example, queries such as \texttt{site:bnpparibas.lu} \texttt{filetype:pdf} allowed us to retrieve comprehensive result lists of publicly available PDF documents. The browser extension \textit{DownThemAll}~\citep{downthemall_repo} was then used to download all the documents that appeared on each result page, ensuring a systematic collection.

\smallskip
\noindent\textbf{Automated Website Crawling.} We used the open-source \textit{WFDownloader App}~\citep{wfdownloader} operating in \textit{Crawler Mode} to systematically collect PDF documents from selected websites. By configuring a crawl depth of 10 and restricting retrieval to PDF files, this method enabled large-scale harvesting of documents from the main websites of the selected banks and their linked subpages.

Through these two strategies, we collected a total of 1,136 documents, all distributed as PDF files, encompassing both standard report-style documents and slide-based presentations.

\subsection{Document Filtering}
In the document filtering stage, the scraped documents are curated through a structured classification process designed to systematically select visually rich documents for subsequent stages of the benchmark construction pipeline. The documents are first sorted by language (English, French, and German). We retain only the English-language documents and remove duplicates, resulting in 791 unique documents for subsequent processing (Table~\ref{tab:en_collected_docs}). These documents are then classified according to their layout characteristics.

The objective of the document classification step is to assign each document to one of two categories: \textit{text-dominant} or \textit{visually rich}. Text-dominant documents consist primarily of continuous text, sometimes complemented by tables. In contrast, visually rich documents rely heavily on layout, tables, charts, or diagrams to convey information that cannot be captured through text alone.
The document classification proceeds as follows. Documents with fewer than 5 pages are discarded, as very short documents are often limited to cover pages, summaries, or annexes and do not provide sufficient structural or visual diversity for reliable classification. For the remaining documents, only the first 20\% of each document is analyzed. This limit strikes a balance between computational efficiency and representativeness, as the early sections of financial reports typically reflect the overall document layout and contain a representative mix of textual and visual elements. Each page in the sampled 20\% of the document is analyzed using a layout detection model trained on PubLayNet~\cite{zhong2019publaynet}, which identifies five layout components: Text, Title, List, Table, and Figure. Pages containing at least one \textit{Table} or \textit{Figure} are labeled as visual. A document is classified as visually rich if visual pages account for at least 50\% of the analyzed subset; otherwise, it is considered text-dominant. 

\subsection{Page Sampling}
The page sampling step selects a representative set of visually rich pages from documents classified as \textit{visually rich} in the previous step, which are then used for Question-Answer (QA) pair generation. Page-level visual richness is determined using the same layout detection model as in document classification: each page is analyzed with a Mask R-CNN model to detect layout elements such as tables and figures, and a page is considered visually rich if it contains at least two \textit{Figure} components.
Sampling is applied at the document level with three constraints: only visually rich pages are eligible, a minimum number of pages is sampled from each document to ensure coverage, and a maximum number is imposed to avoid oversampling very long documents. These constraints limit redundancy and prevent the benchmark from being dominated by pages with similar style, layout or content, thereby preserving diversity. After automatic sampling, a manual verification step removes pages such as cover pages, tables of contents, or pages dominated by photos of people, while a small number of text-dominant pages are intentionally retained to reflect realistic scenarios. In total, 1,000 pages are selected and used in the subsequent Question-Answer generation step.

\subsection{Question-Answer generation}
\label{subsec:qa_generation}
Building on the previously sampled pages, this step automatically generates QA pairs with a particular emphasis on visual elements such as charts and diagrams. Because the objective is to produce questions grounded in visual components, we rely on Vision–Language Models (VLMs), which jointly process text and images, and have recently demonstrated increasingly strong visual understanding capabilities, rather than traditional pipelines that first parse PDFs into text or intermediate formats (e.g., Markdown), followed by the use of Large Language Models (LLMs) for QA generation, as such approaches can easily lead to the loss of visual and layout information.

Multiple prompts and model variants were tested, and after manually inspecting generation quality, GPT-4o~\citep{openai2024gpt4o} was selected as one of the strongest OpenAI models available at the time due to its advanced visual understanding. The prompt used for QA generation (Figure~\ref{fig:prompt_qa_generator}) instructs the model to focus on a single visual element (e.g., table, diagram, or chart), generate reasoning-based questions, and ensure that answers are grounded in the page. Each page yields up to three QA pairs, each consisting of a question, answers, and a hint indicating the answer locations.
\begin{figure}[tbp]
\centering
\begin{tikzpicture}
\tikzset{
  promptbox/.style={
    draw=black,
    fill=gray!6,
    rounded corners=6pt,
    line width=0.4pt,
    inner sep=4pt,
    outer sep=0pt,
    text width=1\linewidth,
    align=left,
    font=\ttfamily\fontsize{6pt}{7pt}\selectfont
  }
}

\node[promptbox] {%
The task is the following:\par
Given a PDF page as an image, you will have to generate questions that can be asked by a user to retrieve information from a large documentary corpus.\par
\vspace{2pt}

Focus on one visual element, such as a table, diagram, image, chart, or any other visually distinct element.\par
Include reasoning or logical deduction.\par
Do not simply use parts of the text as the answer.\par
Do not refer textually to visual elements, for example, ``based on the diagram''.\par
The question must be answerable using the page.\par
The question should be written as if the user does not know that the corpus exists or what it contains.\par
Also generate the answer to the question.\par
The answer must be formatted as a list of words or strings answering the question.\par
Generate at most three Q\&A pairs per page using the following dictionary format, and return only this dictionary:\par
\vspace{2pt}

\{\par
``questions'': [\par
\{ ``question'': ``XXXXXX'', ``answer'': [``YYYYYY''], ``hint'': ``ZZZZZZ'' \}
]\par
\}\par
\vspace{2pt}

NOTE: If there are no questions to ask about the page, return:
\{ ``questions'': [] \}\par
\vspace{2pt}

Here is the page:\par
};

\end{tikzpicture}
\caption{Prompt template used for generating visually grounded question--answer pairs from PDF pages.}
\label{fig:prompt_qa_generator}
\end{figure}

\subsection{Question-Answer review}
To introduce a \emph{human-in-the-loop} component and improve the reliability of the benchmark, we conducted a manual validation step using Label Studio~\citep{Label_Studio} as the annotation platform. The validation was carried out by six doctoral researchers, who reviewed and refined the automatically generated QA pairs through a structured annotation workflow, following a detailed \textit{annotation guideline} that was carefully prepared and provided to all annotators (available in the project repository). The guideline first describes the overall task and validation objectives, then defines the quality dimensions used to assess each QA pair, and finally provides instructions with examples on how to correct or refine QA pairs when needed. 
The following four QA dimensions are evaluated:

\noindent\textbf{Question relevancy}. Whether the question relates to the page content (\emph{relevant} or \emph{irrelevant}).

\smallskip
\noindent\textbf{Answer correctness}. Whether the answer is supported by the page (\emph{correct} or \emph{incorrect}).

\smallskip
\noindent\textbf{Evidence type}. Type of page element supporting the answer: \emph{text}, \emph{table}, \emph{chart}, \emph{diagram}, \emph{image}, or \emph{other}. Multiple evidence types may be assigned to a question.

\smallskip
\noindent\textbf{Question type}. Difficulty is defined by the reasoning required rather than model performance: \textit{Easy} questions involve direct lookup or information extraction, whereas  \textit{Hard} questions require visual interpretation, comparison, or numerical computation.

Following these guidelines, annotators were asked to (i) verify whether the automatically generated questions were appropriate for a RAG setting, in which users typically have no prior knowledge of the underlying document corpus, and (ii) correct or refine questions and answers when necessary by fixing errors or adding missing details. This second step served a dual purpose: aligning questions with realistic RAG scenarios by removing document- or page-specific cues, such as “in page 3” or “in the chart” and improving the factual quality by fixing incorrect answers or adding missing details to the questions to eliminate ambiguity.

Overall, this manual validation process was designed to ensure that the benchmark reflects realistic RAG scenarios, in which users ask questions without knowing whether relevant documents exist or what content they contain, while maintaining high-quality and reliable QA pairs.

\subsection{Query rewriting}
While the Question-Answer review step ensures the correctness, relevance, and factual grounding of QA pairs, the resulting questions may still reflect artifacts of automatic generation or implicit assumptions about document structure. In particular, some questions, although valid, can sound unnatural or overly tailored to the source page, and thus do not fully resemble user queries in realistic RAG settings.
To mitigate this issue, we introduce a \textit{query rewriting} pipeline consisting of two stages. First, the manually validated questions from the previous step are automatically classified according to their suitability for a RAG setting using a GPT-5-based LLM classifier, following the approach of~\citep{wasserman2025real}. Questions judged as \textit{RAG-relevant} are retained without modification, while those identified as not suitable for a RAG setup are forwarded to a rewriting module (see Figure~\ref{fig:benchmark_construction_pipeline}). In the second stage, only the questions classified as not suitable for RAG are reformulated using GPT-5, with the original question and its source page as input, guided by a dedicated VLM-based rewriting prompt (see Figure~\ref{fig:prompt_rewriting}), with the aim of removing document- or page-specific cues while preserving the original semantic intent and reasoning requirements.
\begin{figure}[tbp]
\centering
\begin{tikzpicture}
\tikzset{
  promptbox/.style={
    draw=black,
    fill=gray!6,
    rounded corners=6pt,
    line width=0.4pt,
    inner sep=4pt,
    outer sep=0pt,
    text width=1\linewidth,
    align=left,
    font=\ttfamily\fontsize{6pt}{7pt}\selectfont
  }
}

\node[promptbox] {%
You are given an original query, its source page, and the answer. Your goal is to rewrite each incoming question as a well-formed query for RAG.\par
\vspace{2pt}

Queries that are well-formed for RAG are those that a user might ask without prior knowledge of the document's structure, ensuring they are neither too general nor overly specific to a single page.\par
\vspace{2pt}

Queries that fail this criterion fall into two categories:\par
Queries with explicit page references, such as "in Figure 5" or "the title of the page".\par
Queries that are too broad, such as "What is the net revenue in 2020?" instead of "What is IBM's net revenue in 2020?".\par
\vspace{2pt}

Example of a query not well-formed for RAG:\par
"What is the product roadmap as shown in the diagram on page 21?"\par

\vspace{2pt}

Example of a query well-formed for RAG:\par
"How many active users did Meta's Facebook platform have in Q4 2023?"\par

\vspace{2pt}

Original Query: "\{original\_query\}"\par
Answer: "\{answer\}"\par
Source Page: "\{source\_page\}"\par
};

\end{tikzpicture}
\caption{Prompt used to convert queries into RAG-relevant queries.}
\label{fig:prompt_rewriting}
\end{figure}
The classification and rewriting stages are applied iteratively until all benchmark queries are judged as RAG-relevant. In practice, this process converges quickly, as all benchmark queries were classified as RAG-relevant after two iterations.

\subsection{Hard Negative Page Mining}
To further increase the difficulty of the retrieval task, we augment the benchmark using \textit{hard negative mining}, a process in which, for each query, we select challenging irrelevant pages that are highly similar to the correct page. For each QA pair, we consider only pages originating from the same source PDF as the ground-truth page and retrieve the top-10 pages using the strong textual retriever BAAI/BGE-M3~\citep{bge-m3}. We exclude annotated ground-truth pages from the retrieved results and retain the remaining top-ranked pages as candidate hard negatives. Hard negative pages are then deduplicated across queries before indexing, ensuring that each page appears only once in the final negative pool. This process adds challenging non-relevant pages that are similar in topic and layout to the correct page. This setup evaluates a retriever’s ability to identify the correct page among highly confusable pages, reflecting realistic failure cases in page-level retrieval over heterogeneous document collections.

Overall, our proposed construction pipeline results in a high-quality benchmark of visually grounded QA pairs that reflects realistic Retrieval-Augmented Question Answering scenarios over visually rich financial documents. Summary statistics of the resulting benchmark are reported in Table~\ref{tab:benchmark_stats}, providing an overview of its scale, composition, and diversity.

\begin{table}[!t]
\centering
\caption{Statistics of the \textit{VisR-RAG-Fin} benchmark.}
\small
\setlength{\tabcolsep}{3pt}

\begin{tabular}{@{}c@{\hspace{0.8em}\vrule\hspace{0.8em}}c@{}}
\toprule
\begin{tabular}[t]{@{}l r@{}}
\textbf{Attribute} & \textbf{Count} \\
\midrule
\multicolumn{2}{@{}l}{\emph{Benchmark pages}} \\
\quad Total pages    & 3,377 \\
\quad Positive pages & 802   \\
\quad Negative pages & 2,575 \\
\cmidrule(r){1-2}
\multicolumn{2}{@{}l}{\emph{Questions (by difficulty)}} \\
\quad Total questions & 1,458 \\
\quad Easy questions  & 826   \\
\quad Hard questions  & 632   \\
\end{tabular}
&
\begin{tabular}[t]{@{}l r@{}}
\textbf{Attribute} & \textbf{Count} \\
\midrule
\multicolumn{2}{@{}l}{\emph{Evidence type}} \\
\quad Chart   & 649 \\
\quad Table   & 246 \\
\quad Text    & 435 \\
\quad Diagram & 122 \\
\quad Image   & 35  \\
\quad Other   & 15  \\
\end{tabular}
\\
\bottomrule
\end{tabular}

\label{tab:benchmark_stats}
\end{table}

\section{Experimental Setup}
\raggedbottom
This section details the experimental setup used throughout the paper, including the retrieval and generation configurations, evaluation setup and metrics, tested models, and research questions.

\subsection{Retrieval Configuration}

All retrieval experiments are conducted at the page level over the complete VisR-RAG-Fin page collection. Each retriever indexes pages using its native representation and ranks them using its corresponding scoring function. Sparse textual retrievers use sparse matching scores, dense single-vector retrievers use vector similarity, and multi-vector retrievers use late-interaction scoring. Textual retrievers operate on page content extracted using Unstructured~\cite{unstructured}, including native PDF text and OCR-derived text. Multimodal retrievers directly encode each page as an image. The specific retrieval and context configurations used in each experiment are described in the corresponding research question.

\subsection{Generation Configuration}
Generation uses an end-to-end multimodal RAG pipeline with a fixed prompt and parameters across models. Due to computational constraints, a representative subset comprising two-thirds of the benchmark queries is used for generation experiments. Depending on the experiment, answers are generated either with oracle context, where the ground-truth evidence page is provided, or with retrieval-based context, where top-ranked pages obtained from the retrieval stage are supplied. The modality of the retrieved context varies by experiment and is provided in text, image, or hybrid form according to the retrieval configuration. 

\subsection{Evaluation}
\label{sec:evaluation}
\noindent\textbf{Retrieval.}
For each query, candidate pages are ranked using the retriever's native scoring function, and retrieval effectiveness is evaluated based on the rank of the ground-truth evidence page. We report nDCG@1 and nDCG@3~\citep{jarvelin2002cumulated}, as implemented in the \texttt{ir-measures} library~\citep{DBLP:conf/ecir/MacAvaneyMO22a}, with the analysis focusing on nDCG@3.

\noindent\textbf{Generation.}
Generated answers are evaluated using GPT-5.2 as an LLM judge with a binary correctness decision. Given the original query, ground-truth answer, and generated answer, the judge determines whether the generated answer is fully correct using the prompt shown in Figure~\ref{fig:llm_judge}. Generation is evaluated under two settings. In the oracle setup, answers are generated directly from the ground-truth evidence page, providing an oracle-context reference that isolates generation from retrieval. In the retrieval-based setup, answers are generated only when the ground-truth page appears within the top-$k$ retrieved pages; otherwise, the query is directly counted as incorrect.
\begin{figure}[tbp]
\centering
\begin{tikzpicture}
\tikzset{
  promptbox/.style={
    draw=black,
    fill=gray!6,
    rounded corners=6pt,
    line width=0.4pt,
    inner sep=4pt,
    outer sep=0pt,
    text width=1\linewidth,
    align=left,
    font=\ttfamily\fontsize{6pt}{7pt}\selectfont
  }
}

\node[promptbox] {%
You are an expert QA grader.
Given a question, reference answer, and proposed answer, decide if the proposed answer is fully correct.
Respond ONLY with the single word ``yes'' (fully correct) or ``no'' (anything less than fully correct). No other text.
};

\end{tikzpicture}
\caption{Binary LLM-as-a-judge prompt used for answer correctness evaluation.}
\label{fig:llm_judge}
\end{figure}

\noindent\textbf{LLM judge validation.}
To validate the LLM-based correctness evaluation, we manually annotated a stratified sample of 100 generated answers. The sample was balanced across judge labels and covered the main generation settings, including oracle and retrieval-based setups, different context modalities, top-$k$ values, and generator models. The GPT-5.2 judge achieved 84\% agreement with human labels, corresponding to substantial agreement (Cohen's $\kappa = 0.68$). Disagreements were mostly conservative: 14 human-correct answers were judged incorrect, while only two human-incorrect answers were judged correct.
The same model-independent judging prompt was applied uniformly to all generated answers without revealing the identity of the generator. Although GPT-5.2 was also evaluated as a generator in one experimental condition, the substantial agreement with human annotations supports the overall reliability of the evaluation procedure.

\subsection{Models}
\noindent{\textbf{Retrieval Models.}}
We evaluate a broad range of retriever models spanning both textual and visual modalities. Textual retrievers include dense single-vector models as well as multi-vector late-interaction models, allowing us to assess the impact of representation granularity in text-based retrieval. Multimodal retrievers similarly cover dense vision encoders and multi-vector architectures designed for fine-grained matching over page images. This selection enables a systematic comparison of representation types within and across modalities. Full model names and references are provided in Table~\ref{tab:rq1_results_table}.

\noindent{\textbf{Generation Models.}}
\label{sec:vlms}
We evaluate a set of VLMs used as generators within the evaluated retrieval-augmented pipelines. This panel of VLMs comprises both open-source and closed-source models. The open-source models include InternVL3.5~\citep{wang2025internvl3_5} (2B, 4B, 8B, and 14B variants) and Qwen3-VL~\citep{qwen3vltechnicalreport} (4B and 8B), selected to cover a range of model sizes up to 14 billion parameters and to reflect practical deployment constraints, where fast generation and resource efficiency are critical in real industrial scenarios. Closed-source models are included as strong reference points, namely Gemini-3-Flash~\citep{google2025gemini3flash} and GPT-5.2~\citep{openai2025gpt52}, which represent state-of-the-art proprietary VLMs and provide an upper bound for comparison against the open-source generators. We note that GPT-4o is used only during benchmark construction for initial QA generation, as described in Section~\ref{subsec:qa_generation}, and is not evaluated as an answer-generation model in our experiments.

\subsection{Research Questions}
\label{subsection-research_questions}
We organize our evaluation around the two main components of multimodal RAG: retrieval and generation. RQ1--RQ3 examine retrieval modality, multimodal features, and negative sampling, while RQ4--RQ5 evaluate answer correctness under different context and retrieval settings.

Together, these questions assess both components and their interaction in end-to-end multimodal RAG pipelines:

\noindent\textbf{RQ1 — Retrieval modality.} How do textual and multimodal retrievers compare on page-level retrieval over visually rich financial documents?\par
\noindent\textbf{RQ2 — Retrieval across evidence types.} How do textual and multimodal retrievers perform across text, table, and visual evidence?\par
\noindent\textbf{RQ3 — Impact of negative sampling strategy.} How does retrieval performance change when using hard negatives versus randomly sampled negatives while keeping the candidate pool size fixed?\par
\noindent\textbf{RQ4 — Oracle vs.\ retrieval-based QA.} How does answer correctness change when moving from oracle pages to retrieved pages in an end-to-end multimodal RAG pipeline?\par
\noindent\textbf{RQ5 — Impact of retrieval and context modality.} How do text-only, image-only, and hybrid retrieval-and-context configurations affect answer correctness in a RAG pipeline?

\section{Experimental Results}
\label{sec:experiments}
\newsavebox{\RBbox}
\newenvironment{RQbox}
  {\par\vspace{1pt}%
   \noindent
   \setlength{\fboxsep}{2pt}%
   \begin{lrbox}{\RBbox}%
   \begin{minipage}{\dimexpr\linewidth-2\fboxsep\relax}%
   \color{black}%
  }
  {\end{minipage}%
   \end{lrbox}%
   \colorbox{black!10}{\usebox{\RBbox}}%
   \par\vspace{1pt}%
  }

\definecolor{gain}{RGB}{34,139,34}
\definecolor{loss}{RGB}{178,34,34}

We report results for the research questions defined in Section~\ref{subsection-research_questions}, covering both retrieval and generation components of the multimodal RAG pipeline. Retrieval results assess how well the ground-truth evidence page is ranked among retrieved pages under different design choices, while generation results evaluate answer correctness under oracle and retrieval-based settings.

\begin{RQbox}
\textbf{RQ1 — Retrieval modality.} How do textual and multimodal retrievers compare on page-level retrieval over visually rich financial documents?
\end{RQbox}

\noindent\textbf{Setup.}
We evaluate six textual and twelve multimodal retrievers over the complete VisR-RAG-Fin page collection. The multimodal retrievers comprise five single-vector and seven multi-vector models of different sizes. Retrieval effectiveness is reported using nDCG@1 and nDCG@3.

\noindent\textbf{Results.}
Table~\ref{tab:rq1_results_table} shows that the strongest multi-vector retrievers outperform the strongest single-vector approaches. Among textual models, ColBERTv2 achieves the strongest results in its class, which can be attributed to its late-interaction mechanism that enables fine-grained matching between query and document token representations. This same principle underlies multi-vector multimodal models such as ColPali-1.3 and ColQwen2, contributing to their strong performance when jointly leveraging fine-grained textual cues and spatially grounded visual information at the page level. In contrast, single-vector retrievers, whether textual (e.g., BGE-M3) or multimodal (e.g., GME-Qwen2-2B), compress page content into a single embedding, which limits their ability to capture the heterogeneous and localized signals present in visually rich documents. General-purpose image encoders such as CLIP-L14-336px show weaker performance, as they are trained on broad image--text associations and are not specifically optimized for matching textual queries to document page images. Overall, these results highlight that late-interaction, multi-vector representations are a key factor driving effective page-level retrieval in visually rich financial documents, with the best performance achieved by multimodal models.

From a model size perspective, the trend is not uniform across families. For single-vector multimodal models, scaling shows mixed effects. Increasing Nomic-Embed-MM from 3B to 7B leads to a negligible decrease of 0.09 nDCG@3 (65.89 $\rightarrow$ 65.80), indicating no benefit from additional capacity. In contrast, GME-Qwen2 shows a substantial gain when scaling from 2B to 7B, improving by 13.36 nDCG@3 points (55.29 $\rightarrow$ 68.65). For multi-vector multimodal retrievers, scaling yields consistent improvements within the ColNomic family: increasing the model size from 3B to 7B improves nDCG@3 by 3.37 points (73.67 $\rightarrow$ 77.04), suggesting that larger capacity more effectively enhances late-interaction, multi-vector retrieval. Small multi-vector multimodal retrievers with fewer than 1B parameters can already be highly competitive and even outperform textual retrievers. For instance, ColSmol-256M exceeds the best textual retriever by 19.57 nDCG@3 points, while ColSmol-500M further improves performance and surpasses ColPali-1.3 by 4.36 nDCG@3 points.

\noindent\textbf{Takeaway.}
For page-level retrieval over visually rich financial documents, the strongest late-interaction multimodal retrievers achieve the best performance. Notably, even small multi-vector multimodal models outperform the strongest textual retriever.

\begin{table}[!htbp]
  \centering
  \caption{Retrieval performance measured by nDCG@k ($k=1$ and $k=3$) across textual, single-vector multimodal, and multi-vector multimodal retrievers. Models are grouped by retriever family and size. Best results are shown in bold and second-best are underlined.}
  \small
  \setlength{\tabcolsep}{4pt}
  \renewcommand{\arraystretch}{0.9}
  \begin{tabular}{lcc}
    \toprule
    \textbf{Retriever}
      & \textbf{nDCG@1}
      & \textbf{nDCG@3} \\
    \midrule

    \multicolumn{3}{l}{\textit{Textual retrievers}} \\
    \cmidrule(lr){1-3}
    BM25~\citep{robertson2009probabilistic}        & 19.90 & 28.10 \\
    BGE-M3~\citep{bge-m3}                         & 29.17 & 39.94 \\
    E5-Large~\citep{wang2024multilingual}         & 31.23 & 42.68 \\
    E5-Large-Inst~\citep{wang2024multilingual}    & 29.51 & 41.04 \\
    ColBERTv2~\citep{khattab2020colbert}          & 33.42 & 44.57 \\
    SPLADEv3~\citep{lassance2024spladev3}         & 33.01 & 43.45 \\
    \midrule

    \multicolumn{3}{l}{\textit{Single-vector multimodal retrievers}} \\
    \cmidrule(lr){1-3}
    CLIP-L14-336px~\citep{radford2021learning}            & 18.12 & 26.94 \\
    Nomic-Embed-MM-3B~\citep{nomicembedmultimodal2025}    & 52.30 & 65.89 \\
    Nomic-Embed-MM-7B~\citep{nomicembedmultimodal2025}    & 52.09 & 65.80 \\
    GME-Qwen2-2B~\citep{zhang2024gme}                     & 41.59 & 55.29 \\
    GME-Qwen2-7B~\citep{zhang2024gme}                     & 55.04 & 68.65 \\
    \midrule

    \multicolumn{3}{l}{\textit{Multi-vector multimodal retrievers}} \\
    \cmidrule(lr){1-3}

    \multicolumn{3}{l}{\textbf{Small (less than 1B)}} \\
    \cmidrule(lr){1-3}
    ColSmol-256M~\citep{faysse2024colpali}        & 49.83 & 64.14 \\
    ColSmol-500M~\citep{faysse2024colpali}        & 53.67 & 68.23 \\

    \multicolumn{3}{l}{\textbf{Medium (1B--3B)}} \\
    \cmidrule(lr){1-3}
    ColPali-1.3~\citep{faysse2024colpali}         & 49.90 & 63.87 \\
    ColQwen2~\citep{faysse2024colpali}            & 57.17 & 70.64 \\
    ColQwen2.5~\citep{faysse2024colpali}          & \underline{61.02} & \underline{74.64} \\
    ColNomic-3B~\citep{nomicembedmultimodal2025}  & 60.40 & 73.67 \\

    \multicolumn{3}{l}{\textbf{Large (above 3B)}} \\
    \cmidrule(lr){1-3}
    ColNomic-7B~\citep{nomicembedmultimodal2025}  & \textbf{63.49} & \textbf{77.04} \\
    \bottomrule
  \end{tabular}
  \label{tab:rq1_results_table}
\end{table}

\begin{RQbox}
\textbf{RQ2 — Benefit of multimodal features.} How much does using visual features improve retrieval performance compared to textual retrieval?
\end{RQbox}

\noindent\textbf{Setup.}
We reuse the retrieval results from RQ1 and stratify them by annotated evidence type using three aggregated categories: text, table, and visual, where the visual category groups charts, diagrams, images, and other non-textual elements. Because evidence-type annotations are multi-label, a question may belong to more than one category and is included in each applicable category. We compare the best-performing textual, single-vector multimodal, and multi-vector multimodal retrievers identified in RQ1: ColBERTv2, GME-Qwen2-7B, and ColNomic-7B, respectively. Performance is reported using nDCG@3.

\begin{figure}[!htbp]
    \centering
    \scalebox{0.8}{
        \includegraphics[width=\linewidth]{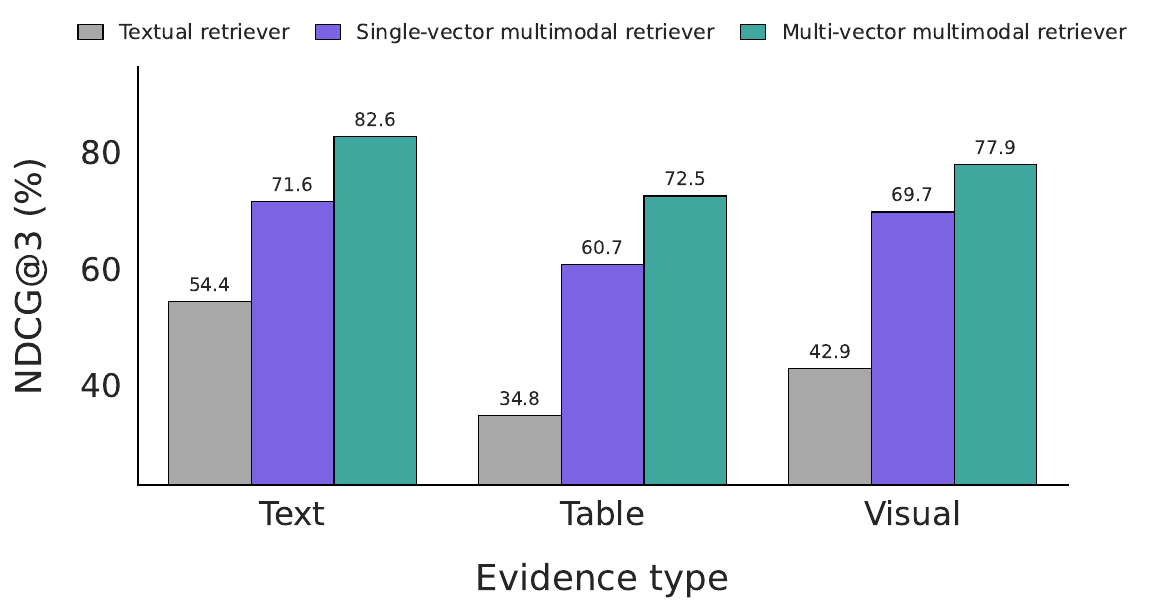}
    }
    \caption{Retrieval performance by evidence type (text, table, visual) comparing the best retriever from each retriever family using nDCG@3.}
    \label{fig:evidence_nDCG3}
\end{figure}

\noindent\textbf{Results.}
Figure~\ref{fig:evidence_nDCG3} shows retrieval performance stratified by evidence type, using the best-performing retriever from each model family as identified in the RQ1 retrieval results (Table~\ref{tab:rq1_results_table}).
Across all evidence types, the multi-vector multimodal retriever achieves the strongest retrieval performance, reaching 82.6 nDCG@3 on text evidence, 72.5 on table evidence, and 77.9 on visual evidence. Compared to the textual retriever, this corresponds to large gains of 28.2 nDCG@3 points on text evidence, 37.7 points on table evidence, and 35.0 points on visual evidence, highlighting the strong performance of multimodal page representations combined with late interaction. The single-vector multimodal retriever also improves substantially over the textual retriever, with gains of 17.2 nDCG@3 points on text evidence, 25.9 points on table evidence, and 26.8 points on visual evidence. However, it remains consistently below the multi-vector multimodal approach, with the largest gap observed on table evidence (72.5 vs.\ 60.7).

The textual retriever relies on extracted page text, including OCR-derived content from visual elements, but this linearized representation does not preserve the spatial structure or localized visual cues needed to match queries to tables, charts, and diagrams. This likely contributes to its lower performance on table (34.8 nDCG@3) and visual (42.9 nDCG@3) evidence compared with the multimodal retrievers, which leverage textual content together with page-level layout and visual information.

\noindent\textbf{Takeaway.}
The evaluated multimodal retrievers outperform the textual retriever across all evidence types. Single-vector multimodal retrieval already provides substantial gains, while the late-interaction, multi-vector multimodal retriever achieves the strongest performance.

\begin{RQbox}
\textbf{RQ3 — Impact of negative sampling strategy.} How does retrieval performance change when using hard negatives versus randomly sampled negatives, while keeping the candidate pool size fixed?
\end{RQbox}

\noindent\textbf{Setup.}
We compare the hard negative benchmark with a random-negative variant containing the same number of candidate pages. Random negatives are uniformly sampled from pages belonging to the same source PDFs as the positive pages. All other experimental conditions are kept unchanged. We evaluate the best-performing textual, single-vector multimodal, and multi-vector multimodal retrievers identified in RQ1 and report nDCG@3.

\noindent\textbf{Results.}
Table~\ref{tab:rq3_negative_sampling} shows that using hard negatives consistently reduces retrieval performance across all retriever families, indicating that hard negative sampling substantially increases task difficulty. Specifically, for the best-performing retriever family, nDCG@3 decreases by 8.92 points for the multi-vector multimodal retriever compared with random negative sampling. Despite this drop, the relative ordering of the selected representative retrievers remains unchanged, with the multimodal retrievers still outperforming the textual retriever. These results show that hard negative mining produces a more challenging retrieval benchmark.

\noindent\textbf{Takeaway.}
Hard negative mining substantially increases benchmark difficulty while preserving the relative ordering of the evaluated retriever families.

\begin{table}[!htbp]
\centering
\caption{Impact of negative sampling strategy on retrieval performance (nDCG@3, scaled to 0–100).}
\scalebox{0.7}{%
\begin{tabular}{llcc}
\toprule
\textbf{Model Type} & \textbf{Negatives} & \textbf{nDCG@3} & $\Delta$ \textbf{nDCG@3} \\
\midrule
\multirow{2}{*}{Textual}
 & Random & 50.69 & --     \\
 & Hard   & 44.57 & $-6.12$ \\
\midrule
\multirow{2}{*}{Single-vector multimodal}
 & Random & 73.76 & --     \\
 & Hard   & 68.65 & $-5.11$ \\
\midrule
\multirow{2}{*}{Multi-vector multimodal}
 & Random & 85.96 & --      \\
 & Hard   & 77.04 & $-8.92$ \\
\bottomrule
\end{tabular}
}
\label{tab:rq3_negative_sampling}
\end{table}

\begin{RQbox}
\textbf{RQ4 — Oracle vs retrieval-based QA.} How does answer correctness change when moving from oracle pages to retrieved pages in an end-to-end multimodal RAG pipeline?
\end{RQbox}

\noindent\textbf{Setup.}
We compare answer correctness under oracle and retrieval-based context. In the oracle setting, the ground-truth page is provided directly to each generator. In the retrieval-based setting, pages are retrieved using ColNomic-7B, the best-performing retriever from RQ1, with $k \in \{1,3,5,10\}$. In both settings, pages are provided as images to the VLMs described in Section~\ref{sec:vlms}, using identical prompts and generation parameters. Answer correctness is evaluated using the binary GPT-5.2 judge described in Section~\ref{sec:evaluation}, and performance is reported as accuracy.

\begin{figure}[!t]
    \centering
    \scalebox{0.9}{%
        \includegraphics[width=\linewidth]{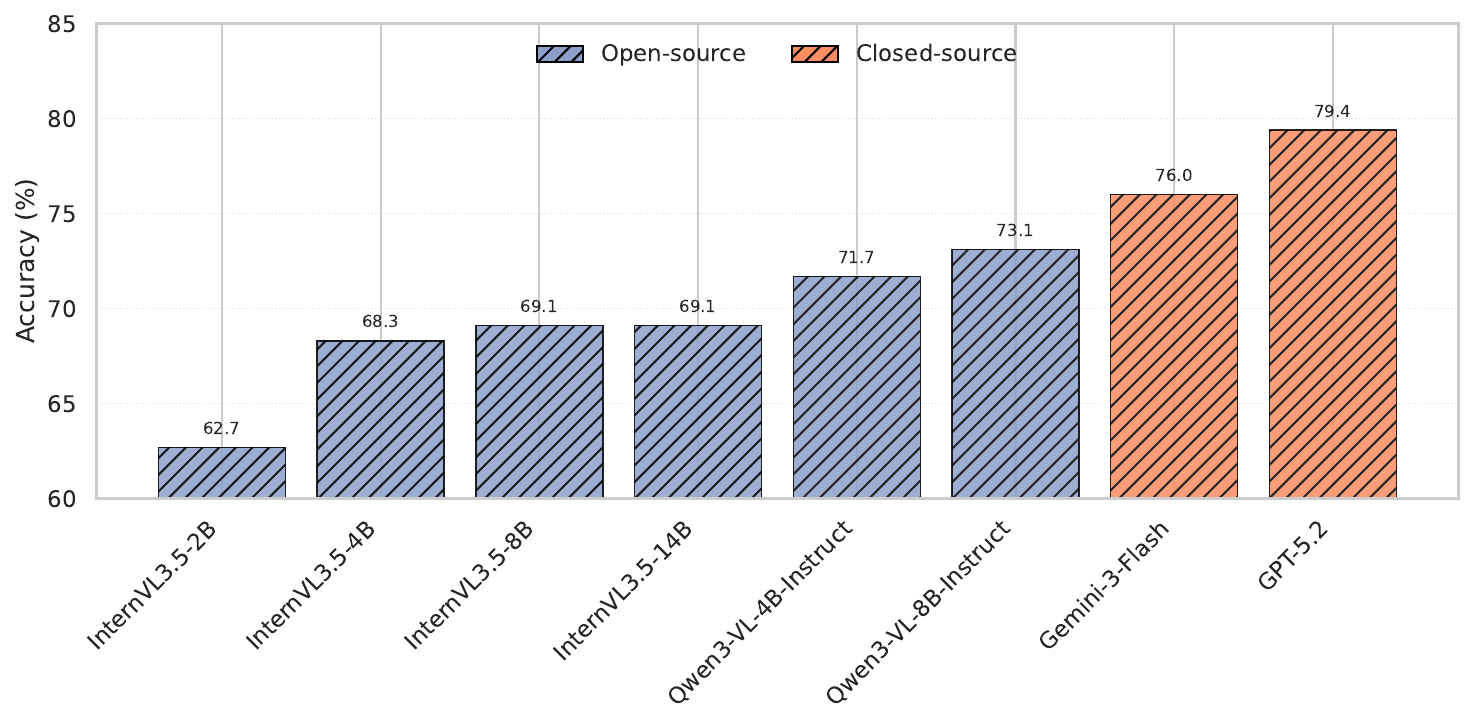}%
    }
    \caption{Oracle accuracy across VLMs. Ordering follows availability (open-source to closed-source) and increasing size within each family. Bar color indicates model availability.}
    \label{fig:oracle_accuracy_plot}
\end{figure}

\noindent\textbf{Results.}
Figure~\ref{fig:oracle_accuracy_plot} shows substantial variation in oracle accuracy across VLMs. InternVL3.5 improves from 62.7\% (2B) to 69.1\% (8B and 14B), while Qwen3-VL reaches 71.7\% (4B) and 73.1\% (8B), outperforming larger InternVL3.5 variants. Closed-source models achieve the highest accuracies, with Gemini-3-Flash reaching 76.0\% and GPT-5.2 reaching 79.4\%. The strongest open-source model, Qwen3-VL-8B, trails these models by 2.9 and 6.3 percentage points, respectively.

\begin{figure*}[!tp]
    \centering
    \scalebox{1}{
        \includegraphics[width=0.90\linewidth]{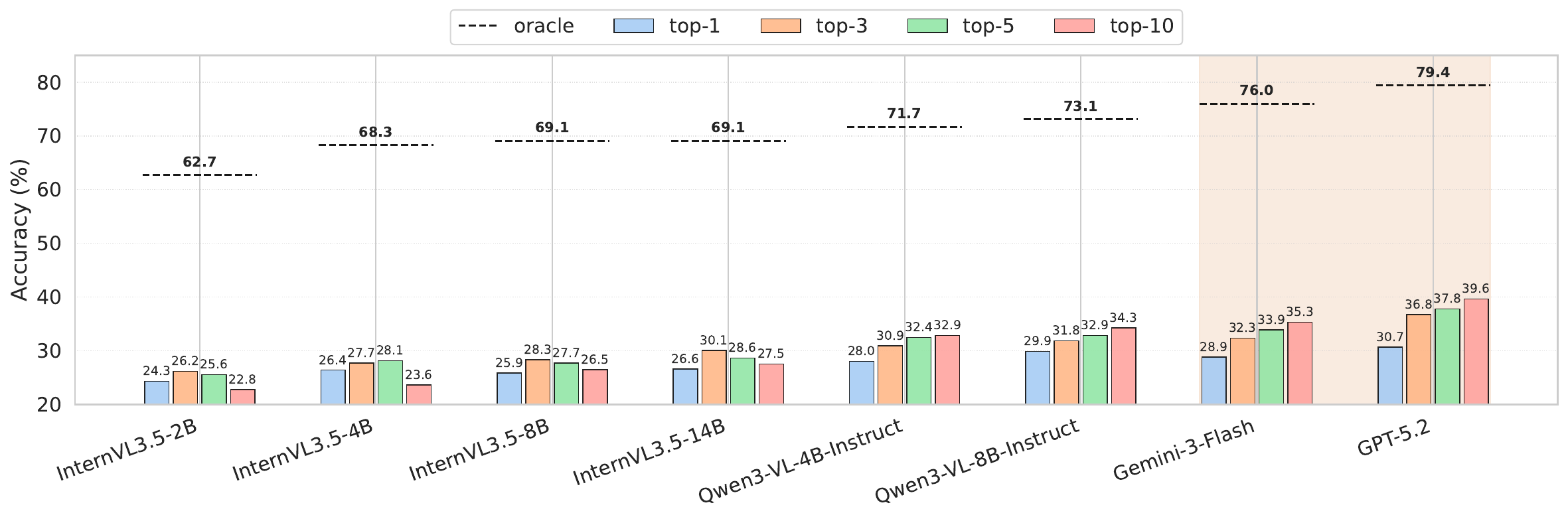}
    }
    \caption{Top-$k$ accuracy across VLMs. Bars report end-to-end accuracy for different retrieval cutoffs $k$, while dashed horizontal lines indicate oracle accuracy (upper bound) for each model. The shaded background highlights closed-source models.}
    \label{fig:topk_accuracy_all_models}
\end{figure*}

As shown in Figure~\ref{fig:topk_accuracy_all_models}, moving from oracle page access to retrieval-based context causes a large drop in answer correctness for every VLM, exposing a substantial gap between oracle and end-to-end RAG settings. Even in the strongest configuration, this gap remains close to 40 points: GPT-5.2 decreases from 79.4\% under oracle access to 39.6\% with top-10 retrieved pages.

Increasing the number of retrieved pages affects model families differently. InternVL3.5 models generally peak at smaller values of $k$ and degrade with additional pages, as illustrated by InternVL3.5-14B decreasing from 30.1\% at top-3 to 27.5\% at top-10. In contrast, Qwen3-VL and closed-source models generally benefit from larger retrieval sets: Qwen3-VL-8B improves from 29.9\% at top-1 to 34.3\% at top-10, while GPT-5.2 improves from 30.7\% to 39.6\%. These trends highlight differences in how VLMs handle additional retrieved context and retrieval noise.

\noindent\textbf{Takeaway.}
VLMs differ substantially in answer correctness even under an oracle context, while the retrieval-based context introduces a large additional performance drop. The ability to exploit additional retrieved pages depends on the model family: some models improve with larger retrieval sets, whereas others become increasingly affected by retrieval noise.

\begin{RQbox}
\textbf{RQ5 — Impact of retrieval and context modality on answer correctness.}
How do text-only, image-only, and hybrid retrieval-and-context configurations affect answer correctness in a RAG pipeline?
\end{RQbox}

\noindent\textbf{Setup.}
We compare three top-10 retrieval-and-context configurations that differ in both the retriever and the modality provided to the generator. In the text-only setting, pages are retrieved using ColBERTv2 and provided as linearized text. In the image-only setting, pages are retrieved using ColNomic-7B and provided as page images. The hybrid setting combines the top-5 textual results as text and the top-5 multimodal results as images, without filtering duplicate pages. Gemini-3-Flash, the second-best-performing VLM in RQ4, is used for answer generation, while GPT-5.2 is used only for evaluation in this experiment. We report global accuracy and accuracy separately for Easy and Hard questions.

\noindent\textbf{Results.}
As noted in the setup, the modality used for retrieval directly determines the modality of the context provided to the generator. Table~\ref{tab:rq5_context_modality} compares answer correctness across the three retrieval-and-context configurations, both globally and by question difficulty. The text-only configuration yields a low global accuracy of 15.7\%, indicating limited effectiveness. In contrast, the image-only configuration reaches 35.3\% global accuracy, outperforming text-only by 19.6 points and achieving the best overall results. The hybrid configuration also delivers large gains over the text-only configuration, reaching 34.1\% accuracy (+18.4 points), but does not surpass the image-only setup.

Importantly, this ranking is preserved when evaluation is stratified by question difficulty. For easy questions, image-only achieves 33.8\% accuracy, compared with 32.5\% for hybrid and 17.4\% for text-only. For hard questions, image-only reaches 37.1\%, followed by hybrid at 36.0\% and text-only at 13.7\%. Overall, configurations that provide visual page images substantially outperform the text-only configuration, while the hybrid setup performs comparably to image-only, with only a small gap between the two.

\noindent\textbf{Takeaway.}
The evaluated image-only and hybrid configurations outperform text-only and achieve comparable accuracy, but this comparison does not isolate the effects of retriever quality and context modality.

\begin{table}[!t]
\caption{End-to-end RAG answer correctness by context modality and question difficulty.
Arrows indicate absolute changes (in percentage points) relative to the text-only configuration.}
\centering
\scalebox{0.80}{
\begin{tabular}{l c cc}
\toprule
\textbf{Modality}
& \textbf{Global Accuracy (\%)}
& \multicolumn{2}{c}{\textbf{Difficulty-Level Accuracy (\%)}} \\
\cmidrule(lr){3-4}
& & \textbf{Easy} & \textbf{Hard} \\
\midrule

Text-only
& 15.7
& 17.4
& 13.7 \\

Image-only
& \textbf{35.3}\,{\scriptsize\color{gain}$\uparrow$\,19.6}
& \textbf{33.8}\,{\scriptsize\color{gain}$\uparrow$\,16.4}
& \textbf{37.1}\,{\scriptsize\color{gain}$\uparrow$\,23.4} \\

Hybrid
& 34.1\,{\scriptsize\color{gain}$\uparrow$\,18.4}
& 32.5\,{\scriptsize\color{gain}$\uparrow$\,15.1}
& 36.0\,{\scriptsize\color{gain}$\uparrow$\,22.3} \\

\bottomrule
\end{tabular}
}
\label{tab:rq5_context_modality}
\end{table}

\section{Limitations}
Our study has several limitations. First, benchmark construction and evaluation rely partly on automated processes, which may introduce residual biases despite human validation. The results also depend on the selected models, prompts, and evaluation settings, while generation experiments use a subset of benchmark queries due to computational constraints.
Second, end-to-end evaluation may conflate retrieval and generation errors, while differences between RAG configurations reflect multiple interacting factors. Text extraction and OCR may also affect textual retrieval performance, particularly for visually structured content.
Finally, VisR-RAG-Fin focuses on visually rich English-language financial documents from a limited set of European banks and uses a controlled retrieval collection. Consequently, the findings may not fully generalize to other domains, languages, or larger enterprise repositories. Fine-grained visual grounding annotations are also outside the benchmark's scope.

\section{Conclusion}
This paper presented a comprehensive study of multimodal Retrieval-Augmented Generation for page-level question answering over visually rich financial documents. We introduced a general benchmark construction pipeline and applied it to build \textit{VisR-RAG-Fin}, an open-source benchmark designed to preserve document layout and visual content while supporting realistic RAG evaluation.

Our empirical analysis leads to several key findings: (1) multi-vector multimodal retrievers consistently provide the strongest retrieval performance,
(2) leveraging visual tokens in multimodal retrieval leads to substantial gains for tables and visually grounded content, (3) as a result, small multimodal retrievers can already outperform the strongest textual retrievers, (4) end-to-end answer correctness drops sharply when moving from oracle to retrieval-based context, revealing retrieval noise as a major bottleneck, and (5) image-only and hybrid RAG configurations outperform the evaluated text-only configuration in answer correctness.

Together, these findings highlight critical design trade-offs in multimodal RAG pipelines. From an industrial perspective, they provide practical guidance for evaluating and building multimodal RAG pipelines over complex, visually rich enterprise documents.





\begin{acks}
This work was supported by BGL BNP Paribas and the Luxembourg National Research Fund (FNR), grant NCER22/IS/16570468/NCER-FT. We thank the BGL DataLab team and the human annotators from the TruX research group at SnT.
\end{acks}




\bibliographystyle{ACM-Reference-Format}
\balance
\bibliography{sample-base}


\end{document}